\documentclass[structabstract]{aa}

\pdfoutput=1
\usepackage[pdftex]{graphicx}
\usepackage{epstopdf}

\usepackage{natbib}
\bibpunct{(}{)}{;}{a}{}{,} 
\usepackage{txfonts}
\usepackage{graphicx}	
\usepackage{amssymb}	
\usepackage{aas_macros}	
\usepackage{subeqn}	
\usepackage{epsfig}	
\usepackage{fixltx2e}	

\newcommand{\beq}{ \begin{eqnarray} }
\newcommand{\eeq}{ \end{eqnarray} }

\newcommand{\yhat}{ {{\bf{\hat y}}} }

\newcommand{\dive }{ {\bf {\nabla}} \cdot }
\newcommand{\curl }{ {\bf {\nabla}} \times }

\newcommand{\boldnabla}{\mbox{\boldmath$\nabla$}}

\def\A{{Alfv\'en }}

\def\ca{{c_A}}

\def\grad{\boldnabla}

\def\muo{\mu_{0}}

\def\yhat{\bf{\hat{y}}}

\def\di{\delta_i}

\date{}			

\begin{document}
\title{Alfv\'en wave phase-mixing and damping in the ion cyclotron range of frequencies}

\author{J. Threlfall\inst{\ref{inst1}}\and K. G. McClements\inst{\ref{inst2}} \and I. De Moortel\inst{\ref{inst1}}}
\institute{School of Mathematics and Statistics, University of St. Andrews, St. Andrews, Fife, KY16 9SS, U.K. \email{jamest@mcs.st-and.ac.uk;ineke@mcs.st-and.ac.uk}\label{inst1}
\and EURATOM/CCFE Fusion Association, Culham Science Centre, Abingdon, Oxfordshire, OX14 3DB, U.K. \email{k.g.mcclements@ccfe.ac.uk}\label{inst2}}

\abstract
{}
{To determine the effect of the Hall term in the generalised Ohm's law on the damping and phase mixing of \A waves in the ion cyclotron range of frequencies in uniform and nonuniform equilibrium plasmas.}
{Wave damping in a uniform plasma is treated analytically, whilst a Lagrangian remap code (Lare2d) is used to study Hall effects on damping and phase mixing in the presence of an equilibrium density gradient.}
{The magnetic energy associated with an initially Gaussian field perturbation in a uniform resistive plasma is shown to decay algebraically at a rate that is unaffected by the Hall term to leading order in $k^2\delta_i^2$ where $k$ is wavenumber and $\delta_i$ is ion skin depth. A similar algebraic decay law applies to whistler perturbations in the limit $k^2\delta_i^2 \gg 1$. In a nonuniform plasma it is found that the spatially-integrated damping rate due to phase mixing is lower in Hall MHD than it is in MHD, but the reduction in the damping rate, which can be attributed to the effects of wave dispersion, tends to zero in both the weak and strong phase mixing limits.}
{}
\keywords{Plasmas - Magnetohydrodynamics (MHD) - Waves - Sun: flares - Sun: chromosphere} 
\maketitle

\section{Introduction}\label{sec:Intro}
The interaction between \A waves and plasma inhomogeneities forms a well-studied and important area of research, for both laboratory and astrophysical plasmas. One process which arises as a result of this interaction is \A wave \emph{phase-mixing}. 
Early studies of \A wave phase-mixing demonstrated a potential for significantly enhanced plasma heating. In particular, \cite{paper:HeyPriest83} proposed the phase-mixing of \A waves as a potential solar coronal heating mechanism through enhanced wave-dissipation.
They outlined that for a magnetohydrodynamic (MHD) treatment of initially planar shear-\A waves, propagating independently on individual magnetic surfaces, large differences in phase are quickly generated between waves on neighbouring field lines, as a result of variation in \A speed across the field. These phase-differences generate progressively smaller scales, and dramatically enhance the effects of viscous and Ohmic dissipation, in the locations where the \A speed gradient is steepest.

This initial concept has been subsequently adapted for a variety of problems using MHD, based on the premise that the waves in question arise as a result of an infinite series of boundary motions at the photosphere. For example, this treatment has been used to investigate the heating of open magnetic field lines under various conditions, e.g. by \citet{paper:Parker1991,paper:IrelandHoodPriest97,paper:deMoortelHoodetal99,paper:DeMoortel2000}. Phase-mixing has also been studied as a source of non-linear coupling to other wave modes \citep[see, e.g.][]{paper:Nakariakov1997,paper:Botha00}.

Of particular interest for this paper is the work of \citet{paper:Hoodetal2002}. They note that an infinite series of boundary motions is unrealistic and instead investigate the effect on the amplitude damping rate due to phase-mixing when the waves are driven by only one or two initial impulsive motions at the boundary. 

Recent studies have also begun to move away from the original MHD treatment, instead focussing on full kinetic descriptions of a plasma undergoing phase-mixing in a collisionless regime, as a potential mechanism for electron acceleration \citep[see, e.g.][]{paper:GenotLouarn04,paper:Tsiklauri05,paper:Tsiklauri08,paper:BianKontar2010}. 
On the assumption that the wavelengths of interest ($\lambda$) are small compared to the particle mean free path $\lambda_{mfp}$, \cite{paper:Tsiklauri05} and \cite{paper:BianKontar2010} model the corona as a collisionless plasma and cite Landau damping as their primary wave dissipation mechanism. Such damping is strongly suppressed if $\lambda \gg \lambda_{mfp}$ \citep{paper:OnoKulsrud1975}, as in the case of propagating EUV disturbances with periods of tens or hundreds of seconds \citep[see e.g.][and references therein]{paper:deMoortel2009}, and a fluid model is then appropriate for the coronal plasma. On the other hand for waves with frequencies approaching the ion cyclotron frequency, typical coronal parameters correspond to classical (Spitzer) collisional mean free paths that exceed $\lambda$, suggesting instead the validity of the collisionless approach. However, \cite{paper:CraigLitvinenko2002} have shown that it is not appropriate to use classical resistivity under flaring conditions because it implies current scale lengths that are several orders of magnitude shorter than $\lambda_{mfp}$, and moreover is inconsistent with the electric fields required to account for the observed acceleration of protons to tens of MeV on timescales of the order of one second \citep{paper:Hamilton2003}. \citeauthor{paper:CraigLitvinenko2002} proposed that the effective resistivity under flaring conditions (specifically in a reconnecting current sheet) is determined by turbulence arising from electron-ion drift (e.g. ion acoustic) instabilities, and deduced that this effective resistivity could exceed the classical value by a factor of around $10^6$. Under these circumstances a fluid model can be appropriate even for relatively high frequency waves. This may also be true in the upper chromosphere, where the plasma is both cooler and denser (and consequently much more collisional) than in the corona.

It is well known that the electron inertia term in the generalised Ohm's law becomes comparable to the MHD terms when the system lengthscale approaches the electron skin depth, $\delta_e$, which in a low beta plasma can exceed the ion Larmor radius. Moreover for perturbations with frequencies approaching the ion cyclotron frequency $\Omega_i$, the Hall term in the generalised Ohm's law becomes as important as the MHD terms when the lengthscale of the system approaches the ion skin depth $\delta_i$ ($\gg \delta_e$). When the introduction of the Hall term into Ohm's law is the only modification made to the otherwise standard set of MHD equations, we may refer to this as a Hall MHD system.

Hall MHD has been found to be important for a number of fundamental plasma processes. For example, in magnetic reconnection studies, \citet{paper:Birn2001} found that all models which included Hall dynamics returned indistinguishable reconnection rates, concluding that the inclusion of the Hall term is the minimum requirement for fast reconnection (for a summary of Geospace Environmental Modelling (GEM) challenge results, see \cite{book:BirnPriest}).

High frequency waves (i.e. with frequencies $\omega\sim\Omega_i$), have been observed in a range of astrophysical plasma systems, for example in the solar corona (summarised in \cite{review:Marsh2006}) and in situ, at the Earth's bow shock \citep{paper:Sckopke1990}. When oscillations in this frequency range are excited in collision-dominated plasmas, such that the collisional mean free path is less than the wavelength, it is then appropriate to use a Hall MHD model.

The goal of our work is to determine the extent to which the main consequences of phase-mixing (wave dissipation and plasma heating) are affected solely by the addition of the Hall term to Ohm's law. To do this, we first investigate the damping rate of a uniform plasma using Hall MHD (Section \ref{sec:analysis}). Phase-mixing is then included, by allowing the
equilibrium density to vary (Section \ref{sec:nonUniform}). Numerical simulations of a Hall MHD system, with various density profiles and Hall term strengths, are described in Section \ref{sec:results}. We interpret these results and present conclusions in Section \ref{sec:discussion}.

\section{Wave Damping Analysis}\label{sec:analysis}
The Hall MHD form of Ohm's Law is
\[
{\bf{E}}+{\bf{v}}\times{\bf{B}}={\eta}{\bf{j}}+\frac{1}{ne}{\bf{j}\times{\bf{B}}},
\]
where ${\bf{E}}$ and $\bf{B}$ are electric and magnetic fields, ${\bf{v}}$ is the single-fluid plasma velocity, ${\bf{j}}$ is current density, $\eta$ is electrical resistivity, $n$ is plasma number density and $-e$ is electron charge. Linearising the induction equation corresponding to this form of Ohm's law, together with the equation of motion (neglecting the pressure gradient force), for a plasma with a uniform equilibrium field ${\bf{B_0}}=B_0\yhat$ for a constant $B_0$, uniform number density and resistivity, and zero equilibrium flow, gives:
\begin{eqnarray}
\frac{\partial{\bf{B_1}}}{\partial t}&=&\curl\left({{\bf{v}}\times{\bf{B_0}}}\right)+\frac{\eta}{\muo}\boldnabla^2{\bf{B_1}}
-\frac{1}{\muo ne}\curl\left[\left(\curl{\bf{B_1}}\right)\times{\bf{B_0}} \right] \label{eq:HallInduct_L} \\
\frac{\partial{\bf{v}}}{\partial{t}}&=&\frac{1}{{\rho_0}\muo}\left(\curl{\bf{B_1}}\right)\times{\bf{B_0}}, \label{eq:EOM_L}
\end{eqnarray} 
where ${\bf{B_1}}[=(B_x,0,B_z)]$ now represents a transverse perturbation to the equilibrium field, $\rho_0$ is equilibrium mass density, and $\muo$ is the permeability of free space. Note that although, for simplicity, we neglect plasma pressure, by assuming $\beta=0$ in this analytical treatment (as Alfv\'en, whistler and ion cyclotron waves are all incompressible in a linear regime), the numerical simulations presented in Section \ref{sec:results} incorporate plasma pressure, i.e. $\beta\neq0$.\\
By inserting (\ref{eq:EOM_L}) into the time derivative of (\ref{eq:HallInduct_L}) in the usual manner, the linearised \A wave equation is then modified:
\begin{eqnarray}
 \frac{\partial^2 B_c}{\partial t^2}= \ca^2\frac{\partial^2 B_c}{\partial y^2} +\left( \frac{\eta}{\muo}-i\ca\delta_i\right) \frac{\partial^2}{\partial y^2}\left( \frac{\partial B_c}{\partial t}\right),
\label{eq:wave1}
\end{eqnarray}
where we have expressed transverse field perturbations in the form of a complex variable $B_c=B_z+iB_x$, and introduced the \A speed, $\ca=B_0/\sqrt{\rho_0\muo}$, and the ion skin depth, $\delta_i=c/\omega_{pi}=c\sqrt{m_i\epsilon_0/ne^2}$, where $m_i$ is the ion mass, $c$ is the speed of light, and $\epsilon_0$ is the permittivity of free space.\\
Seeking wave-like solutions of the form $\exp{\left(i\left[ky-\omega t\right]\right)}$ allows us to form a dispersion relation to express perturbation frequencies, $\omega$, as a function of wavenumber $k$. By considering only the regime of weak damping ($\eta^2k^2/\muo^2\ca^2\ll1$), we obtain two separate solutions depending on the size of the parameter $k^2\di^2$. Taking first the case of $k^2\di^2\ll1$, we make use of a simple Taylor expansion to find (to leading order in $k^2\di^2$):
\begin{equation}
 \omega =\pm\ca k\pm\ca\di\frac{k^2}{2}-\frac{i\eta k^2}{2\muo},   
\label{eq:smallkdiomega}
\end{equation}
which describes a shear \A wave, modified by Hall effects, and subject to resistive damping.\\
Considering the opposite case, when $k^2\di^2\gg1$, we find:
\[
 \omega=\pm \ca\di\frac{k^2}{2}\pm\ca\di \frac{k^2}{2}\sqrt{1+\frac{2i\eta}{\muo\ca\di}+\frac{4}{k^2\di^2}}-\frac{i\eta k^2}{2\muo}. 
\]
Including the next term in the Taylor expansion, we find two distinct forward propagating solutions:
\begin{subequations}
 \begin{equation}
 \omega_{+}=\frac{\ca}{\di}-\frac{i\eta}{\muo\di^2}, 
 \label{eq:highkdiomega+}
 \end{equation}
 \begin{equation}
 \omega_{-}=\ca\di k^2-\frac{i\eta k^2}{\muo},
 \label{eq:highkdiomega-} 
\end{equation}
\label{eq:highkdi}
\end{subequations}
where in this limit, we now obtain a combination of whistler and ion cyclotron (i.c.) waves, both subject to a form of resistive damping.

\subsection{Long Wavelength Hall MHD Regime (Uniform)}\label{subsec:kdill1}
We can examine the effect of this difference in behaviour of both $k^2\di^2$ regimes, by focussing on the evolution of an initially Gaussian pulse (of width $\sigma$, and amplitude $B_i$), which is allowed to travel along the equilibrium field, taking the form:
\begin{equation}
 B_c(y,0)=B_z+iB_x=B_i\exp{\left(-\frac{y^2}{2\sigma^2} \right) }.
\label{eq:initial_pulse}
\end{equation}
Our complex variable can be interpreted as a Fourier integral, evolving in time as:
\begin{equation}
 B_c=B_c(y,t)=\int^{\infty}_{-\infty} f\left(k\right)\exp{\left\lbrace i\left(ky-\omega t\right)\right\rbrace} dk,
\label{eq:Bc}
\end{equation}
with $f(k)$ determined by the initial conditions:
\begin{equation}
 f(k)=\frac{1}{4\pi}\int^{\infty}_{-\infty}B_c(y,0)\exp\left(iky\right)dy=
\frac{B_i\sigma}{2\sqrt{2\pi}}\exp{\left(-\frac{k^2\sigma^2}{2}\right) },
\label{eq:f(k)}
\end{equation}
where we have used the standard result \citep[Eq. 7.4.6]{book:AbramStegun}:
\begin{equation}
 \int^{\infty}_{0}\exp\left(-\beta x^2\right)\cos\left(b x\right) dx=\frac{1}{2}\sqrt{\frac{\pi}{\beta}}\exp\left(-\frac{b^2}{4\beta} \right).
\label{eq:AbramStegun} 
\end{equation}
We also use this result to evaluate the integral in Eq. (\ref{eq:Bc}) in the limit $k^2\di^2\ll1$, finding:
\begin{equation}
 B_c=\sum_{+,-}\frac{B_i}{{\sqrt{1+\left(\frac{\eta}{\muo}-i\ca\di \right)\frac{t}{\sigma^2} }}} {\exp{\left( \frac{-\left(y\pm\ca t \right)^2}{2\left[\sigma^2+\left(\frac{\eta}{\muo}-i\ca\di\right)t\right] }\right) }}
\label{eq:B_c(y,t)}
\end{equation}
where the summation is over forward- and backward-propagating waves. Eq. (\ref{eq:B_c(y,t)}) describes a pair of pulses travelling in opposite direction at approximately the \A speed, which are damped by finite resistivity and circularly polarised. 
We can also calculate the contribution to the total energy per unit area in ($x$,$z$), ${\cal E}_{B_{TOT}}$, made by the magnetic energy
per unit area in ($x$,$z$) of both pulses (${\cal E}_{B_c}$) as follows:
\[
 {\cal E}_{B_{TOT}}=\frac{1}{2\muo}\int^{\infty}_{-\infty} \left( {\bf{B_0}}+{\bf{B_1}}\right)^2 dy=\frac{B_0^2}{2\muo}+{\cal E}_{B_c}.
\]
Since the magnetic perturbation is transverse to the equilibrium field, ${\bf B_0}\cdot{\bf B_1} = 0$ and hence
\begin{equation}
{\cal E}_{B_c}=\frac{1}{2\muo}\int {\bf{B_1}}^2 dy = \frac{1}{2\muo}\int^\infty_{-\infty} B_cB_c^{*} dy.
\label{eq:EBc}
\end{equation}
Many of the factors in $B_cB_c^{*}$ will cancel upon integration, hence the energy associated with the pulse (${\cal E}_{B_c}$) evolves as:
\begin{equation}
 {\cal E}_{B_c}^{k^2\di^2\ll1}=\frac{\sqrt{\pi}\sigma B_i^2} {4\muo{\left( {1+{\eta t}/{\muo\sigma^2} }\right)^{1/2} }}\left\lbrace 1 + {\exp{\left( -\frac{\ca^2 t^2}{\sigma^2+{\eta t}/{\muo}}\right) }} \right\rbrace.
\label{eq:EBclowlimit}
\end{equation}
Thus after a short initial transient phase (essentially the time taken for an \A wave to travel a distance equal to the initial pulse width, $\sigma$), we recover a power law decay ($\propto t^{-1/2}$ for $t\gg\muo\sigma^2/\eta$) in the energy associated with the pulse. This expression (Eq. \ref{eq:EBclowlimit}) is compared with several numerical simulation results in Section \ref{sec:results}, and can be seen in Fig. \ref{fig:enmatch}. It is straightforward to show that the expression given by Eq. (\ref{eq:EBclowlimit}) for the Hall MHD long wavelength ($k^2\di^2\ll1$) regime is \emph{identical} to that found for an initially Gaussian pulse in the MHD limit.

\subsection{Short Wavelength Hall MHD Regime (Uniform)}\label{subsec:kdigg1}
Turning to the opposite limit, $k^2\di^2\gg1$, the perturbation frequencies (\ref{eq:highkdi}) comprise of a combination of resistively damped whistler and ion cyclotron waves. We can again describe the pulse evolution associated with each separate wave branch in this limit, in the manner described previously (Section \ref{subsec:kdill1}), again for an initially Gaussian perturbation. Beginning with (\ref{eq:Bc}), and with the same initial conditions (\ref{eq:f(k)}) as the previous limit, we find that the whistler wave calculation proceeds similarly to that of the previous section, however the i.c. wave (being independent of wavenumber) differs somewhat:
\begin{subequations}
 \begin{equation}
  B_c^{w}=\frac{B_i\sigma}{\sqrt{2\pi}}\int^{\infty}_{0} 
e^{iky}\exp{\left( -\frac{\sigma^2 k^2}{2}-\frac{\eta k^2 t}{\muo}-i\ca\di k^2 t\right) } dk,
\label{eq:Bcw1}
 \end{equation}
 \begin{equation}
 B_c^{ic}=\frac{B_i\sigma}{\sqrt{2\pi}}\exp{\left(-\frac{ i\ca t}{\di}-\frac{\eta t}{\muo\di^2}  \right) }  \int^{\infty}_{0}e^{iky}\exp{\left( \frac{-k^2\sigma^2}{2}\right) }dk.
\label{eq:Bcic1}
 \end{equation}
\label{eq:Bclargeint}
\end{subequations}
Evaluating the integrals in Eq. (\ref{eq:Bclargeint}), using Eq. (\ref{eq:AbramStegun}), we obtain:
\begin{subequations}
 \begin{equation}
  B_c^{w}=\frac{B_i}{{\sqrt{1+2\left(\frac{\eta}{\muo}-i\ca\di \right)\frac{t}{\sigma^2} }}} {\exp{\left( \frac{-y^2}{2\left[\sigma^2+2\left(\frac{\eta}{\muo}-i\ca\di\right)t\right] }\right) }}
\label{eq:Bcw2}
 \end{equation}
 \begin{equation}
 B_c^{ic}=B_i\exp{\left(-\frac{ i\ca t}{\di}-\frac{\eta t}{\muo\di^2}  \right) }
\exp{\left(-\frac{y^2}{2\sigma^2} \right) }.
\label{eq:Bcic2}
 \end{equation}
\label{eq:Bclarge}
\end{subequations}
In this short wavelength ($k^2\di^2\gg1$) regime, the peak of the pulse now no longer propagates, but decreases in amplitude. The right circularly polarised component of the pulse rapidly broadens, due to the high whistler speed, and damps algebraically at a rate similar to that found in both the MHD and long wavelength ($k^2\delta_i^2 \ll 1$) Hall MHD regimes. The left circularly polarised (ion cyclotron wave) component, on the other hand, damps exponentially. It should be noted that this damping arises from resistive dissipation, and as such should be distinguished from the kinetic ion cyclotron damping arising from wave-particle interactions.

We may, again, calculate the energy associated with each solution (\ref{eq:Bclarge}), using (\ref{eq:EBc}), where we still only obtain transverse perturbations, and hence ${\bf{B_1}}\cdot{\bf{B_0}}$ still makes no contribution to the energy. In this limit, we obtain an expression for the energy associated with the individual whistler and i.c. wave branches:
\begin{subequations}
 \begin{equation}
{\cal E}_{B_c}^{w}=\frac{B_i^2\sigma\sqrt{\pi}}{2\muo}\left(1+\frac{2\eta t}{\muo\sigma^2} \right)^{\bf{-1/2}}, 
\label{eq:EBcw}
 \end{equation}
 \begin{equation}
{\cal E}_{B_c}^{ic}=\frac{B_i^2\sigma\sqrt{\pi}}{2\muo}\exp{\left(-\frac{2\eta t}{\muo\di^2}\right) }.
\label{eq:EBcic}
 \end{equation}
\label{eq:EBclarge}
\end{subequations}
Thus, for waves in the $k^2\di^2\gg1$ regime, we no longer see the initial transient phase seen previously in (\ref{eq:EBclowlimit}), and for long timescales the algebraically-damped whistler contribution to the wave energy is dominant over the exponentially-damped contribution from the ion cyclotron wave.

\subsection{Wave Damping and Phase-Mixing in a Non-Uniform Plasma}\label{sec:nonUniform}
By now allowing the equilibrium plasma density to vary in a direction perpendicular to both the direction of the equilibrium field ($y$) and the direction of initial perturbation ($z$), we can investigate what effect the Hall term has on the dissipation rate in a non-uniform plasma. When the gradients in the $x$-direction are sufficiently large, and the effects of viscosity are negligible, the linearised \A wave equation in the MHD limit takes the form [\cite{paper:Hoodetal2002}]:
\begin{eqnarray}
 \frac{\partial^2 B_z}{\partial t^2}= \ca^2(x)\frac{\partial^2 B_z}{\partial y^2} + \frac{\eta}{\muo} \frac{\partial^2}{\partial x^2}\left( \frac{\partial B_z}{\partial t}\right).
\label{eq:wave2}
\end{eqnarray}
The variation in \A speed $\ca(x)=B_0/\sqrt{\muo\rho(x)}$ causes steep gradients to build up in the direction of the inhomogeneity which, in turn, significantly enhances resistive damping in the regions where the inhomogeneity is greatest. \citet{paper:Hoodetal2002} used a multiple time-scale analysis to derive from Eq. (\ref{eq:wave2}) a one-dimensional diffusion equation whose solutions can be expressed in terms of the \A speed gradient $\ca'(x)=d\ca(x)/dx$. For the case of the initially Gaussian pulse defined by Eq. (\ref{eq:initial_pulse}), the forward-propagating solution takes the form:
\begin{equation}
 B_z=\frac{B_i}{2\sqrt{1+\ca'^2\eta t^3/3\muo\sigma^2}}\exp{\left(-\frac{\left(y-\ca t\right)^2 }{2\left[\sigma^2+\ca'^2\eta t^3/3\muo\right] }\right) }.
\label{eq:Hood1}
\end{equation}
We can evaluate the perturbed magnetic field energy per unit length in the z-direction ${\cal E}_{B_c}^H$ for this case by integrating $B_z^2/2\mu_0$ over a finite distance $x_0< x < x_1$ in the $x$-direction and from minus to plus infinity in the
y-direction, taking into account the presence of both forward- and backward-propagating pulses. The y-integration can be performed analytically, yielding:
\begin{eqnarray}
 {\cal E}_{B_c}^{\mathrm H}&=&\frac{B_i^2\sigma\sqrt{\pi}}{8\muo}\int_{x_0}^{x_1}\frac{dx}{\left( 1+\ca'^2\eta t^3/3\muo\sigma^2\right)^{1/2} } \nonumber \\
&+&\frac{B_i^2\sigma\sqrt{\pi}}{8\muo}\int_{x_0}^{x_1}\frac{\exp{\left(-\frac{\ca^2 t^2}{\sigma^2+\ca'^2\eta t^3/3\muo} \right) }}{\left( 1+\ca'^2\eta t^3/3\muo\sigma^2\right)^{1/2} }dx.
\label{eq:Hood2pulseenergy}
\end{eqnarray}
The integrals in this expression can be readily evaluated numerically, to allow comparison with our numerical simulations in the next Section.

\section{Simulation Results}\label{sec:results}
The system was modelled numerically using a two dimensional version of a Lagrangian remap scheme ({\tt{LareXd}}), described by \cite{paper:LareXd01}, which includes an optional Hall physics package to incorporate the Hall term into the standard MHD system of equations, seen here in normalised dimensionless form: 
\begin{eqnarray*}
&&\frac{\partial \rho}{\partial t}+\dive{\left(\rho{\bf{v}} \right) }=0, \\
&&\rho\left( \frac{\partial {\bf{v}}}{\partial t}+\left( {\bf{v}}\cdot\boldnabla\right){\bf{v}} \right)= \left(\curl{\bf{B}}\right)\times{\bf{B}}-\grad{p}, \\
&&\frac{\partial {\bf{B}}}{\partial t}=\curl{\left( {\bf{v}}\times{\bf{B}}\right) }-\curl{\left(\eta\curl{\bf{B}} \right) } -\lambda_i\curl{\left[\frac{1}{\rho}\left(\curl{\bf{B}}\times{\bf{B}} \right)  \right] }, \\
&&\rho\left( \frac{\partial\epsilon}{\partial t}+\left( {\bf{v}}\cdot{\boldnabla}\right)\epsilon\right)  =-p\dive{\bf{v}}+\eta j^2,
\end{eqnarray*}
for dimensionless mass density $\rho$, pressure $p$, magnetic field strength ${\bf{B}}$, fluid velocity ${\bf{v}}$, internal energy density $\epsilon$, resistivity $\eta$ (the reciprocal of the Lundquist number) and ion skin depth $\lambda_i(=\di/l_0)$.

In Section \ref{sec:Intro}, two particular regimes were identified where collisional models, such as this, may be more appropriate to describe the plasma behaviour than collisionless treatments. Normalising temperatures ($T_0$) and densities ($n_0$) using flaring coronal values ($T_0\approx2\times10^{6}{\mbox{K}}$ and $n_0\approx10^{16}{\mbox{m}^{-3}}$) or upper chromospheric values ($T_0\approx2\times10^{4}{\mbox{K}}$ and $n_0 \approx 2\times10^{16}{\mbox{m}^{-3}}$) places the simulations firmly within these regimes. Specifying a low plasma beta ($\beta=0.01$) in the simulations fixes the normalising magnetic field strengths to $B_0\approx118\mbox{G}$ in the flaring corona, or $B_0\approx17\mbox{G}$ in the upper chromosphere. This also determines the effective size of several fundamental plasma parameters, outlined for these normalising values in Table \ref{tab:lengths}.
\begin{table}[ht]
	\caption{Approximate Lengthscales}
     \label{tab:lengths}
     \centering
          \begin{tabular}{ccc}\hline\hline
	Parameter	& Flaring Corona & Chromosphere \\
              \hline
	 electron gyro-radius ($r_e$)	& $0.26\mbox{cm}$ 	& $0.18\mbox{cm}$ \\
	 ion gyro-radius ($r_i$)	& $11\mbox{cm}$ 	& $7.9\mbox{cm}$ \\
	electron skin depth ($\delta_e$)& $5.3\mbox{cm}$		& $3.8\mbox{cm}$ \\
	ion skin depth ($\delta_i$) 	& $2.3\mbox{m}$ 	& $1.6\mbox{m}$ \\
classical mean free path ($\lambda_{mfp}$)& $30\mbox{km}$ 	& $1.5\mbox{m}$ \\
              \hline
          \end{tabular}
\tablefoot{Approximate values of fundamental plasma parameters, calculated using normalisation values. In the flaring corona, these were density $n\sim1\times10^{16}\mbox{m}^{-3}$, temperature $T\sim2\times10^6\mbox{K}$ and magnetic field strength $B\sim118\mbox{G}$, while in the chromosphere, values of $n\sim2\times10^{16}\mbox{m}^{-3}$, $T\sim2\times10^4\mbox{K}$ and $B\sim17\mbox{G}$ were used.}
\end{table}

The parameter $\lambda_i$ (which controls the effect of the Hall term in our simulations) was initially set equal to $0.0072$. Using this value of $\lambda_i$, together with the ion skin depths listed in Table \ref{tab:lengths}, implies a normalising lengthscale $l_0=0.3{\mbox{km}}$ in the flaring corona, and $l_0=0.2{\mbox{km}}$ in the chromosphere. Given that the mean square wavenumber for a Gaussian pulse of width $\sigma$ can be found using $\langle k^2\rangle=1/\sigma^2$, a choice of width $\sigma=0.1$ places the simulations firmly within the long wavelength regime (discussed in Section \ref{subsec:kdill1}), as $\langle k^2\rangle \di^2\approx0.005\ll1$. By relating simulated perturbation frequencies ($\omega$) to the perturbation size (by assuming $\omega\sim k\ca$), our choices for $\lambda_i$ and $\sigma$ perturb the simulations with frequencies which have begun to approach the ion cyclotron frequency ($\omega/\Omega_i\sim k\di\sim\lambda_i/\sigma\sim0.07$).

As pointed out by \citet{paper:CraigLitvinenko2002}, the resistivity could be as much as a factor of $10^6$ higher than the classical Spitzer value in the flaring corona. Using this enhancement factor and the normalisation described above for flaring conditions, we find $\eta=0.0005$. Note that using the same value for $\eta$ and the chromospheric normalisation, the enhancement of the resistivity as compared to the Spitzer value, is less than $10^2$. An enhancement in the resistivity implies a
corresponding reduction of the mean free paths quoted in Table \ref{tab:lengths}. These reduced mean free paths are then much smaller than the typical wavelengths of the modes under consideration, so that Hall MHD is an appropriate model to use.

In order to study phase-mixing, we allow the equilibrium density to vary along $x$, with the form: 
\begin{equation}
 \rho(x)=\frac{1}{\left(1-\alpha+\alpha\cos{\left(2\pi x\right) }\right) ^{1/2}},
\label{eq:rho}
\end{equation}
chosen for constant density at the edges, $\rho(0)=\rho(1)=1$, and a central increase in density controlled by a steepness parameter, $\alpha$. We also vary the specific internal energy density of the system, $\epsilon$, allowing us to define a constant plasma pressure in terms of the plasma beta, $\beta(=0.01)$, and the magnetic pressure, as:
\[
 \epsilon(x)= \frac{\beta |B|^2}{2\rho(x)\left(\gamma-1\right) }, \quad p=\rho(x)\epsilon(x)\left(\gamma-1\right)=\frac{\beta|B|^2}{2}.
\]
We set up a constant equilibrium field to reflect the analytical setup, with ${\bf{B_0}}(=[{B_{0x}},{B_{0y}},{B_{0z}}])=[0,1,0]$, which we perturb in the form outlined by Eq. (\ref{eq:initial_pulse}). As our analytical work is based on a linearised system of equations, we fix the pulse amplitude, $B_i(=0.0005)$ to be small enough that we minimise non-linear effects. 
A $2000\times2000$ grid was used to fully resolve the effects of a density gradient in $x$ and potential whistler/i.c. effects in $y$. As the density and energy density profiles are symmetric, periodic boundary conditions are used in $x$, whilst the $y$ boundaries are left open, and the simulation terminated before the pulse reaches the boundaries. The range of the numerical box (set, for the $k^2\di^2\ll1$ simulations, at $-10\leq y\leq10$, $0\leq x\leq1$) has a large range in $y$, in anticipation of the pulse propagation. 

\begin{figure}[ht]
 \centering
\resizebox{\hsize}{!}{\includegraphics{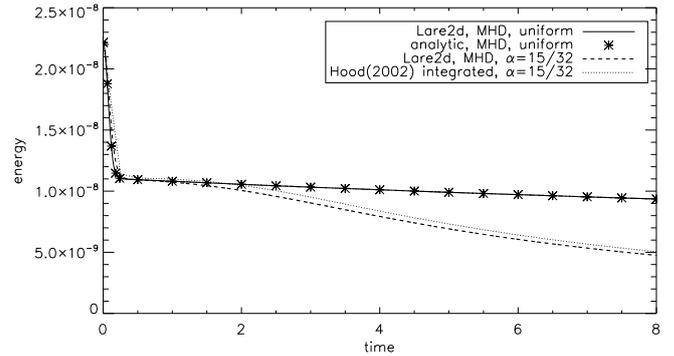}} 
 \caption{Temporal evolution of magnetic energy perturbation associated with initially Gaussian pulse, computed analytically for cases with uniform equilibrium density using Eq. (\ref{eq:EBc}) (stars), and strong equilibrium density gradient using Eq. (\ref{eq:Hood2pulseenergy}) (dotted curve). The solid and dashed curves show the corresponding numerical results, obtained using {\tt{Lare2d}}.}
 \label{fig:enmatch}
\end{figure}

Our initial investigation sought to recover, using {\tt{Lare2d}}, the behaviour of the pulse and its evolution along the equilibrium field, in the long wavelength Hall MHD regime (expressed in Eq. \ref{eq:B_c(y,t)}). For this investigation, selecting a steepness factor of $\alpha=0$ in our equilibrium density expression (Eq. \ref{eq:rho}) provided the required uniform equilibrium density and uniform internal energy in the simulations. We found that the evolution of the pulse in the numerical scheme exactly matched the equivalent analytical expression (Eq. \ref{eq:B_c(y,t)}), both in the MHD and long-wavelength Hall MHD ($\lambda_i=0.0072$) limits.
\begin{table}[ht]
	\caption{Density Enhancement Values}
     \label{tab:alpha}
     \centering
          \begin{tabular}{cc}\hline\hline
	      Steepness Parameter ($\alpha$) & $\rho_{max}-\rho_{min}$ \\
              \hline
              $1/10$ & $0.12$ \\
              $9/50$ & $0.25$ \\
              $5/18$ & $0.50$ \\
              $3/8$ & $1.00$ \\
              $4/9$ & $2.00$ \\
              $15/32$ & $3.00 $ \\
              \hline
          \end{tabular}
\tablefoot{Values of $\alpha$ are chosen to provide specific sizes of density enhancement, $\rho_{max}-\rho_{min}$, at the centre ($x=0.5$) using Eq. (\ref{eq:rho}).}
\end{table}
We also sought to recover the expression for the evolution of perturbed magnetic energy associated with the pulse in the long wavelength Hall MHD regime (Eq. \ref{eq:EBclowlimit}) numerically. This too showed excellent agreement with the derived expression, and a comparison of the numerical and analytical energies can be seen in Fig. \ref{fig:enmatch}.

Moving to the non-uniform equilibrium density simulations, we then selected a range of specific equilibrium density enhancements through the steepness parameter $\alpha$ according to Eq. (\ref{eq:rho}). The specific values of $\alpha$ chosen, and the corresponding density enhancements are listed in Table \ref{tab:alpha}.
\begin{figure}[htb]
 \centering 
\resizebox{\hsize}{!}{\includegraphics{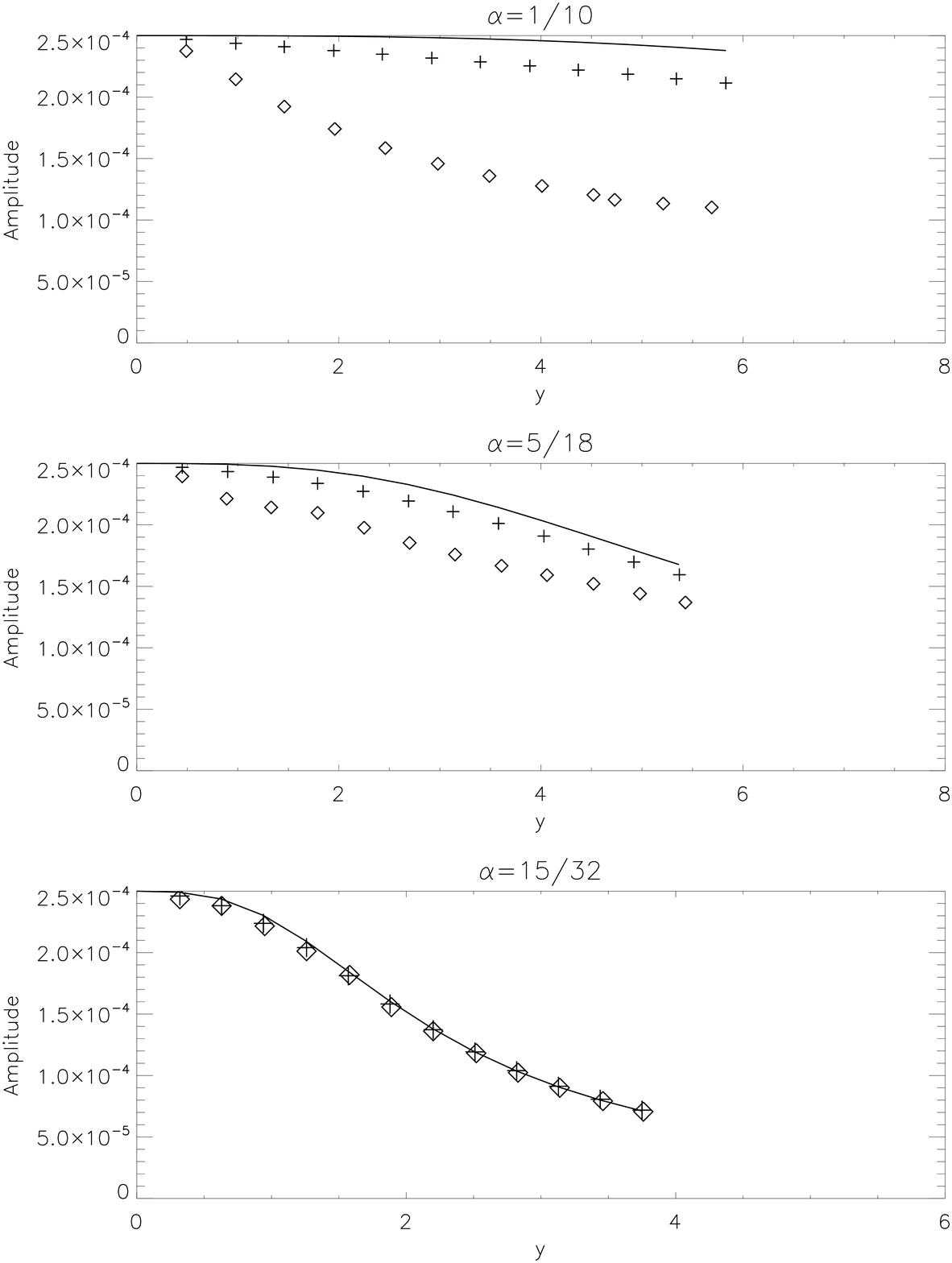}}
\caption{Comparison of predicted amplitude fall-off for an analytical treatment of MHD phase-mixing (Eq \ref{eq:Hood1} - solid curve), with tracked pulse peak of {\tt{Lare2d}} simulations.Displayed are the results of the MHD (crosses) and long wavelength Hall MHD (diamonds) simulations for three different density gradients (for corresponding density enhancements, see Table \ref{tab:alpha}).} 
\label{fig:ampmatch}
\end{figure}

As discussed previously, \citet{paper:Hoodetal2002} obtained an expression describing the evolution of a pulse as it undergoes phase mixing in the MHD limit [Eq.(\ref{eq:Hood1})]. We sought to test the validity of this expression for a range of equilibrium density gradients using {\tt{Lare2d}}, by tracking the height and location of the peak of the pulse as it travelled along the equilibrium field, and comparing this with the pulse amplitude damping rate found by \citet{paper:Hoodetal2002}. We found that the MHD simulations with large equilibrium density increases displayed good agreement with the proposed damping rate, but this agreement was lessened by reducing the equilibrium density enhancement. Furthermore, we repeated the simulations for the long wavelength Hall MHD regime, again finding that the amplitude evolution of the pulse was in agreement with the expression proposed by \citet{paper:Hoodetal2002}, but only in the cases where the density enhancement was sufficiently large. These results are summarised in Fig. \ref{fig:ampmatch}.

We also evaluated numerically the integrals in Eq. (\ref{eq:Hood2pulseenergy}) to investigate how the magnetic energy associated with a pulse, which undergoes MHD phase-mixing, evolves in time (see Section \ref{sec:nonUniform}). As we have shown, the expression of \citet{paper:Hoodetal2002} displays good agreement with the simulations in the cases of steepest density gradient. We therefore selected the steepest of our density gradient cases ($\alpha=15/32$), prescribing a particular equilibrium density $\rho(x)$ (from Eq. \ref{eq:rho}), and hence obtained an expression for the gradient in \A speed, $\ca'(x)$, required for the numerical integration of Eq. (\ref{eq:Hood2pulseenergy}). Using {\tt{Lare2d}}, we then compared this with the evolution of magnetic energy of an identical pulse and equilibrium density gradient, also using MHD. This comparison can be seen in Fig. \ref{fig:enmatch}, together with the magnetic energy evolution of an identical pulse for the uniform density case, and the analytical expression derived for the uniform case, both in MHD. In the uniform density case there is a very close agreement between the analytical and {\tt{Lare2d}} results. When a density gradient is present, the field perturbation energy starts to decay slightly earlier in the {\tt{Lare2d}} simulations than it does in the analytical result. In all cases, there is a short transient phase after $t=0$ in which the field perturbation energy drops to half its initial value and the flow energy reaches approximate equipartition with the field energy. As discussed in Section \ref{sec:analysis}, the duration of this transit phase is essentially determined by the \A propagation time across the initial pulse width.

\begin{figure*}[hbtp]
 \centering 
\resizebox{\hsize}{!}{\includegraphics{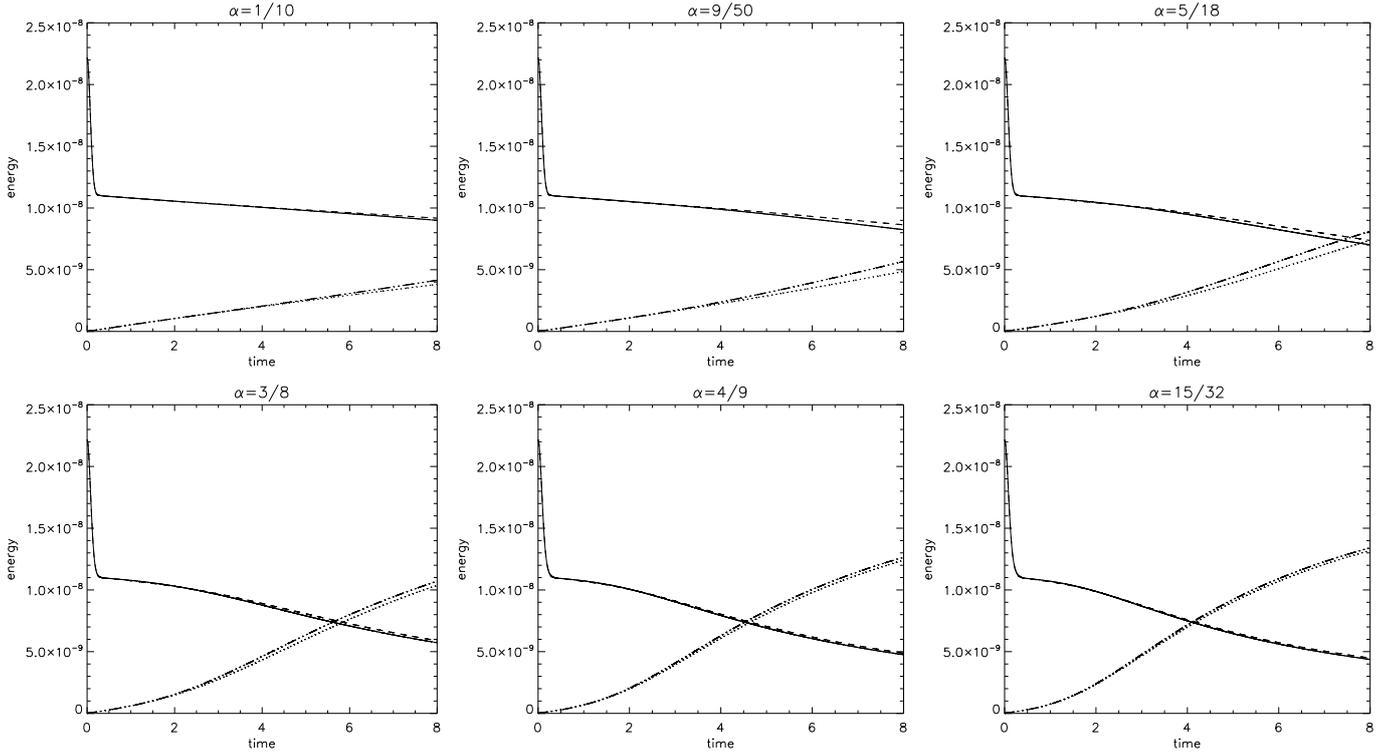}}
  \caption{Evolution of field perturbation energy in MHD (solid curve) and $\lambda_i=0.0072$ Hall MHD (dashed curve) simulations for six values of density steepness parameter $\alpha$. The internal energy is also plotted for the MHD (dot-dash curve) and $\lambda_i=0.0072$ Hall MHD (dotted curve) simulations.}
 \label{fig:megaplot2}
\end{figure*}

Having studied the behaviour of the perturbed magnetic energy evolution in the uniform and steep density gradient limits using MHD simulations, we then investigated the energy response for a variety of different density gradients, and at different skin depths. For clarity, we also included the perturbed internal energy evolution of the pulse. Our first set of results compare the evolution of the pulse in MHD, with that of the $\lambda_i=0.0072$ simulations, corresponding to the long wavelength Hall MHD regime, discussed in Section \ref{subsec:kdill1}. These results are seen in Fig. \ref{fig:megaplot2}.

\begin{figure}[hbtp]
 \centering 
\resizebox{\hsize}{!}{\includegraphics{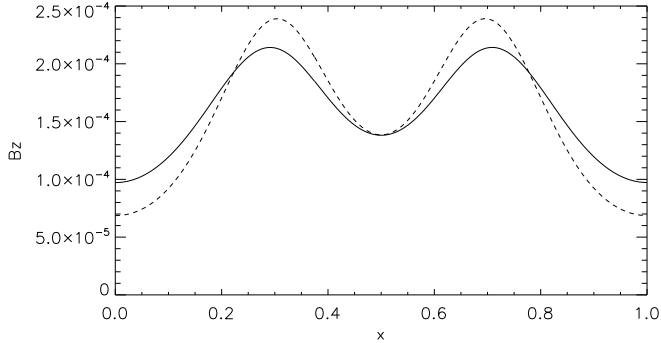}}
  \caption{Comparison of two snapshots of slices (in x) through the pulse amplitude, at location (in y) of maximum phase-mixing. Snapshots shown are for long wavelength Hall MHD (solid) and MHD (dashed) cases, taken at the same time, for an identical initial pulse and with density steepness $\alpha=5/18$.}
 \label{fig:slice}
\end{figure}

Figure \ref{fig:megaplot2} clearly displays differences between the MHD and Hall MHD cases, for density gradients which are neither uniform nor very steep. To gain insight into the reasons for these differences, we have also tracked the evolution of the pulse profile in the $x$-direction as it travels at the \A speed of the steepest density region. In the Hall MHD cases, we observe consistently smaller amplitude gradients. As an example of this, Fig. \ref{fig:slice} shows snapshots of MHD and Hall MHD pulse profiles in simulations with $\alpha=5/18$. One might have expected the dispersive effects of the Hall term to induce a cascade into short wavelengths, which could in principle invalidate our neglect of kinetic effects. However, Fig. \ref{fig:slice} shows that the opposite effect occurs: the Hall term actually reduces the gradients resulting from phase mixing.

\begin{figure*}[hbtp]
 \centering 
\resizebox{\hsize}{!}{\includegraphics{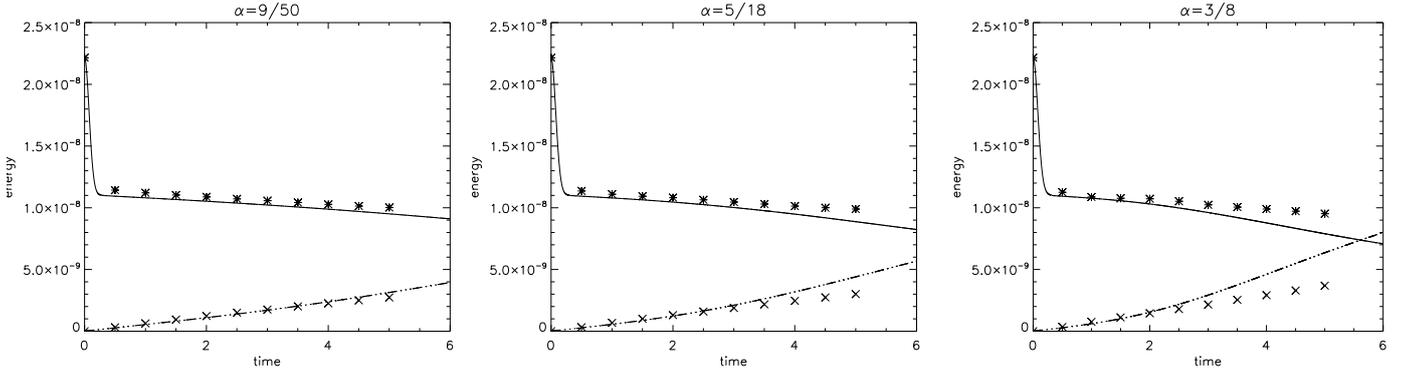}}
  \caption{Evolution of field perturbation energy in MHD (solid curve) and $\lambda_i=0.072$ Hall MHD (stars) simulations for three intermediate values of density gradient parameter $\alpha$. Also plotted is the internal energy for the MHD (dot-dash curve) and $\lambda_i=0.072$ Hall MHD (crosses) simulations.}
 \label{fig:megaplot3}
\end{figure*}

We have demonstrated analytically that the plasma response dramatically differs in the short wavelength ($k^2\di^2\gg1$) Hall MHD limit, from that of the MHD and long wavelength Hall MHD limits (Section \ref{sec:analysis}). Our final investigation sought to begin to bring out the behaviour of this limit by increasing $\lambda_i$ by a factor of 10. In doing so, we also
reduce the associated enhancement factor in resistivity required to justify setting $\eta = 0.0005$ in the code. The enhancement factor is reduced to $10^5$ for the flaring corona, whilst for the upper chromosphere the effective resistivity is less than an order of magnitude larger than the equivalent Spitzer value, using the values outlined earlier in Section \ref{sec:results}. In these simulations, we also increased the size of the numerical box (to $-30\leq y\leq30$, $0\leq x\leq1$, accounting for larger dispersive effects and the faster whistler wave component) and the number of gridpoints (now at $6000\times2000$, maintaining the same resolution). The perturbed frequencies in these simulations are now a large fraction of the ion cyclotron frequency ($\omega/\Omega_i\sim\lambda_i/\sigma\sim0.7 $).
Fig. \ref{fig:megaplot3} compares the response of the perturbed magnetic and internal energies in simulations of a Hall MHD plasma with $\lambda_i=0.072$, with that of an MHD plasma, for three density gradient cases which are neither uniform, nor dramatically varying. It is clear that in this Hall MHD regime, there is a strong reduction in the damping rate compared to the MHD limit.

\section{Discussion and Conclusions}\label{sec:discussion}
The behaviour demonstrated by the analysis of a uniform plasma in the long wavelength Hall MHD regime (Section \ref{subsec:kdill1}) was fully recovered by the {\tt{Lare2d}} simulations. Both the simulated pulse evolution and its associated magnetic energy are indistinguishable from the expressions derived analytically (Eq. (\ref{eq:B_c(y,t)}) and Eq. (\ref{eq:EBc}) respectively). The simulations show that the MHD results remain linearly polarised, whilst the long wavelength Hall MHD pulse becomes circularly polarised. However, both the simulations and the analytical results tell us that this has no impact on the magnetic energy associated with the pulse, which damps at the same rate in both regimes.

Considering now the results of the investigation concerning a non-uniform equilibrium density, Fig. \ref{fig:ampmatch} shows that the amplitude damping rate described by \citet{paper:Hoodetal2002} is well matched to the rate of damping of the peak of an MHD pulse when the gradient in \A speed is sufficiently steep. Whilst not a particularly surprising result, as the strong phase mixing limit required for Eq. (\ref{eq:wave2}) to be valid applies in this case, it is somewhat more surprising to find for sufficiently steep \A speed gradients, that the long wavelength Hall MHD regime also conforms to this amplitude damping rate. For shallower gradients in \A speed, Fig. \ref{fig:ampmatch} also demonstrates that this damping rate quickly becomes inappropriate for the $k^2\di^2\ll1$ Hall MHD simulations, whilst giving better agreement with the MHD evolution of the peak of the pulse. This is partly due to the fact that in the Hall regime, the pulse is no longer linearly polarised, quickly losing amplitude to the other field components. The Hall regime is also subject to significant dispersive effects, which reduce the amplitude of the pulse as it spreads along the equilibrium field. However, for sufficiently steep density gradients, phase-mixing becomes the dominant cause of amplitude dissipation.

By comparing the energy evolution associated with our numerical simulations of an MHD pulse to that found by numerically integrating an expression for the pulse evolution given by \citet{paper:Hoodetal2002} in Fig. \ref{fig:enmatch}, we see that, apart from a slight initial delay, the two results appear closely matched. The reason for the difference between the two is simply due to the treatments used. The linearised \A wave equation (\ref{eq:wave2}) used in the treatments of \citet{paper:HeyPriest83} and \citet{paper:Hoodetal2002} retains only the $x$-derivatives, on the basis that derivatives along the field are small compared to those across it. However, in our simulations, the (initially uniform) pulse propagates a finite distance in which the $y$-derivative causes resistive damping along the equilibrium field. This continues until the gradients in $x$ build up sufficiently to allow phase-mixing to become the dominant damping mechanism. The brief initial period of longitudinal damping is not included in the treatment of \citet{paper:Hoodetal2002}, resulting in the discrepancies in behaviour seen between the energy evolution of the two non-uniform scenarios shown in Fig. \ref{fig:enmatch}.

Our final investigation sought to compare the response of the magnetic and internal energies associated with the pulse for different equilibrium density gradients and at different skin-depths. Fig. \ref{fig:megaplot2} shows that for shallow density
gradients, both MHD and long wavelength Hall MHD simulations display similar behaviour to that seen in the uniform case. For the steepest density gradients considered, again the long wavelength Hall MHD and MHD simulations exhibit similar behaviour, suggesting that MHD phase-mixing dominates the energy evolution (see Fig. \ref{fig:enmatch}). However, we see significant departures from the MHD behaviour of both perturbed magnetic and internal energies associated with the long wavelength Hall MHD regime, in the cases where the density is neither uniform nor sharply varying. In these cases, the inclusion of the Hall term causes the damping rate of perturbed magnetic energy to be reduced. Consequently there is a slower increase in the internal energy of the pulse than that seen in MHD: phase-mixing of the plasma causes less rapid plasma heating in the Hall MHD regime than in the MHDregime.

Evidence to support this can be seen in Fig. \ref{fig:slice}, where we plot a slice (in $x$) through pulse amplitude at the location (along $y$) of maximum phase-mixing, both for the $\lambda_i=0.0072$ and MHD simulations. The figure clearly demonstrates that in the Hall case, the gradients in $x$ are smaller than those in the MHD case. The whistler component of the Hall MHD pulse displays highly dispersive behaviour, which increases with skin depth. This dispersion spreads the pulse envelope along the equilibrium field as it travels, introducing higher wavenumbers into the pulse but only in the equilibrium field direction. In the density gradient direction, the group dispersion of the wave has the effect of reducing the damping rate by making amplitude gradients across the field (in $x$) smaller, reducing the efficiency of phase-mixing and its damping of the magnetic energy of the pulse.

This behaviour is also seen in the higher skin depth Hall MHD simulations. In the $\lambda_i=0.072$ simulation results, seen in Fig. \ref{fig:megaplot3}, there is now a very significant reduction in the damping rate of the magnetic energy, even in the steep density gradient case, which, in the long wavelength Hall MHD regime has begun to converge to the MHD results. As the skin depth is increased, the equilibrium density gradient must also be increased in order to recover a rate of energy dissipation comparable to those seen in the long wavelength Hall MHD and MHD limits.

In contrast to the results presented here, \citet{paper:Tsiklauri05} and \citet{paper:BianKontar2010} found that by extending the MHD treatment of phase mixing to include kinetic effects it was possible to demonstrate an {\it{enhanced}} rate of wave dissipation rather than a reduced rate. Specifically, \citet{paper:Tsiklauri05} observed \A wave damping in a collisionless particle-in-cell simulation of a plasma with an equilibrium density gradient, the amplitude decay law being similar to that found by \citet{paper:HeyPriest83} using an MHD treatment. \citet{paper:BianKontar2010} used drift-kinetic theory (applicable to mode frequencies $\omega \ll \Omega_i$) to argue that this result can be attributed to the generation of a parallel electric field $E_{\parallel}$ associated with the presence of finite perpendicular wavenumbers and consequent mode conversion to kinetic \A waves; because $E_{\parallel}$ is finite the wave energy can be dissipated via Landau damping. The results of our fluid study show that extensions of phase mixing theory to include non-MHD effects do not necessarily lead to an enhanced damping rate in the case of waves with $\omega \sim \Omega_i$ and in the presence of collisions. In view of the differences between the physics assumptions of Hall MHD on the one hand and those of collisionless Vlasov or drift-kinetic theory on the other, it is not
particularly surprising that the two models produce different results with regard to phase mixing. The collisionless regime is the relevant one for high frequency waves in the corona when the resistivity is close to classical. Our calculation, on the other hand, may be more applicable to either flaring conditions in the corona or the upper chromosphere. In our current study, we have only included dissipation through resistivity, neglecting other dissipative mechanisms. \cite{paper:CraigLitvinenko2005} noted that classical viscosity could play a significant role in flare energy release. Hence, a possible extension of the work reported here would be to examine the effects of the Hall term on phase mixing when viscous terms are added to the momentum and energy equations.

In summary, we have described analytically the propagation of an initially Gaussian field perturbation along a uniform equilibrium field in the presence of resistivity, and the evolution of magnetic energy associated with this perturbation. While the evolution of the perturbation differs in the MHD, long wavelength and short wavelength Hall MHD regimes, the energy evolution is the same in the MHD and long wavelength Hall MHD cases. In a non-uniform equilibrium plasma, our simulations show that the damping rate for the energy associated with the pulse in Hall MHD is significantly reduced when the \A speed variation is neither uniform nor sharply varying, compared to that of an MHD treatment as the Hall term actually reduces the gradients resulting from phase-mixing. Moreover, as the ion skin depth is increased, the density gradient needed for MHD phase-mixing to dominate the evolution of the pulse (the "strong phase-mixing limit") must also increase.

\begin{acknowledgements}
This work was funded by the United Kingdom Engineering and Physical Sciences Research Council, under grant EP/G003955 and a CASE studentship, and by the European Communities under the contract of Association between EURATOM and CCFE. The views and opinions expressed herein do not necessarily reflect those of the European Commission. We thank P. J. Cargill \& T. Neukirch (University of St Andrews) for helpful discussions. IDM acknowledges support of a Royal Society University Research Fellowship.
\end{acknowledgements}

\bibliographystyle{aa}        
\bibliography{15479}          
\end{document}